\documentclass[12pt,a4paper]{report}

\usepackage{graphicx} 
\usepackage{fancyhdr} 
\usepackage[utf8]{inputenc}
\usepackage[T1]{fontenc} 
\usepackage{setspace} 
\usepackage{mathptmx} 
\usepackage{slantsc}
\usepackage{titlesec}
\usepackage{mfirstuc} 
\usepackage{calc}
\usepackage[acronym, nonumberlist]{glossaries} 
\usepackage[defernumbers=true, sorting=none]{biblatex}
\usepackage{hyperref} 
\usepackage{pdfpages}
\usepackage{float}
\usepackage{minitoc}
\usepackage{pdflscape}
\usepackage{booktabs}
\usepackage{caption}
\usepackage{subcaption}

\fancypagestyle{preliminary}{
    
    \fancyhead[RCL]{}

    \pagenumbering{Roman}
}
\fancypagestyle{chapter}{
    \fancyhead[L]{\normalfont\itshape\fontsize{10pt}{12pt}\selectfont\nouppercase{\leftmark}}
    \fancyhead[R]{}

    \fancyfoot[C]{\thepage}

    \pagenumbering{arabic}
}

\newcommand{\headingBaseline}{12}
\newcommand{\headingBaselineDiv}{10}
\newlength{\chapterFontSize}
\newlength{\sectionFontSize}
\newlength{\subsectionFontSize}
\newlength{\chapterBaseline}
\newlength{\sectionBaseline}
\newlength{\subsectionBaseline}

\titleformat{\chapter}[display]
    {\normalfont\bfseries\fontsize{\chapterFontSize}{\chapterBaseline}\selectfont}{\chaptertitlename\ \thechapter}{14pt}{}
\titlespacing{\chapter}{0pt}{10pt}{25pt}

\titleformat{\section}[hang]
    {\normalfont\bfseries\fontsize{\sectionFontSize}{\sectionBaseline}\selectfont}{\thesection}{5pt}{}
\titlespacing{\section}{0pt}{25pt}{15pt}

\titleformat{\subsection}[hang]
    {\normalfont\bfseries\itshape\fontsize{\subsectionFontSize}{\subsectionBaseline}\selectfont}{\thesubsection}{5pt}{}
\titlespacing{\subsection}{0pt}{20pt}{10pt}

\makenoidxglossaries

\DeclareBibliographyCategory{cited}
\AtEveryCitekey{\addtocategory{cited}{\thefield{entrykey}}}
\hypersetup{
    colorlinks = true,
    linkcolor = blue, 
    citecolor = blue, 
    urlcolor = blue, 
    filecolor = black 
}

\newcommand{\thesisTitle}{Design of a Dual-Band Patch Antenna with stacked structure}

\newcommand{\authorName}{Xunchi Ma}
   
\newcommand{\degreeQualification}{Bachelor of Engineering in Telecommunications Engineering}

\newcommand{\school}{School of Engineering and Physical Sciences}
\newcommand{\university}{Heriot-Watt University}
\newcommand{\monthDate}{April}
\newcommand{\yearDate}{2025}
\newacronym{fss}{FSS}{ Frequency Selective Surface}
\newacronym{ibfd}{IBFD}{in-band full-duplex}
\newacronym{wban}{WBAN}{wireless body area network}
\newacronym{ism}{ISM}{industrial, scientific, and medical}
\newacronym {cst}{CST}{Computer Simulation Technology Studio Suite}
\usepackage{subfiles}

\begin{document}

\dominitoc

\pagestyle{empty}
\begin{center}
\vspace*{15pt}\par
\setstretch{1}
\begin{spacing}{1.8}
{\Large\bfseries\MakeLowercase{\capitalisewords{\thesisTitle}}}\\
\end{spacing}

\vspace{40pt}\par
\includegraphics[width=140pt]{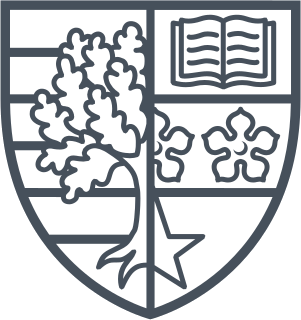}\\
\vspace{40pt}\par

{\itshape\fontsize{15.5pt}{19pt}\selectfont by\\}\vspace{15pt}\par

{
\Large \authorName
}\vspace{55pt}\par

{
\large An Honours report submitted for the Degree of \\ \vspace{8pt} \Large\slshape\degreeQualification\\
}

\vspace{35pt}\par

{\scshape\setstretch{1.5} \school\\ \university\\
}

\vspace{50pt}\par

{\large \monthDate\ \yearDate}

\vfill

\begin{flushleft}
\setstretch{1.4}\small
The copyright in this thesis is owned by the author. Any quotation from the thesis or use of any of the information contained in it must acknowledge this thesis as the source of the quotation or information.
\end{flushleft}
\end{center}

\clearpage
\pagestyle{preliminary}
\begin{center}
\LARGE\textbf {Abstract}
\end{center}
\vspace{5pt}

\noindent This report presents the design, simulation, and fabrication of a compact, dual-band patch antenna using a stacked structure, developed as part of an undergraduate engineering research project. Patch antennas are widely valued in modern wireless communication systems for their low profile, light weight, and ease of integration, but they traditionally suffer from limited bandwidth and single-frequency operation. In response to the increasing demand for antennas capable of operating at multiple frequency bands within a compact form factor, this project explores a dual-band design targeting the 2.25–2.35 GHz and 5.6–5.8 GHz frequency ranges—commonly used in industrial, scientific, and medical (ISM) applications.

\noindent Two design approaches were considered: a slotted patch and a stacked configuration. After simulation-based analysis, the stacked structure was selected for its superior performance. The final design features two radiating copper layers separated by dielectric substrates, forming a multilayered structure where the top patch is responsible for higher-frequency resonance and the middle patch supports the lower band. A single feed line connects to the top layer, simplifying the design and minimizing the overall thickness of the antenna. To fine-tune performance, a 5 mm displacement between patch centers was introduced, which improved electromagnetic coupling and achieved the desired dual-band behavior.

\noindent Simulation results demonstrated promising performance, with strong return loss (S11) values and directional gain in both frequency bands, indicating good impedance matching and radiation characteristics. However, due to time constraints and limited access to professional measurement equipment such as a vector network analyzer (VNA) or anechoic chamber, physical validation was limited. Despite these challenges, the final prototype achieved functionality aligning with design goals and highlights the potential of stacked patch antennas for applications in compact wireless devices, wearable technologies, and embedded systems. This project serves as a foundation for further development and optimization of dual-band antenna systems in practical wireless communication scenarios.
\clearpage
\begin{center}
\LARGE\textbf {Acknowledgements}
\end{center}
\vspace{5pt}

\noindent I would like to express my sincere gratitude to my final year project supervisor, Dr. Dimitris Anagnostou, whose expert guidance, unwavering support, and insightful feedback have been invaluable throughout this journey. His dedication and encouragement have greatly enriched my academic experience and inspired me to strive for excellence.

\noindent I am also deeply thankful to Adán Simón for his technical assistance and to all the staff at the Electronic Workshop for fabricating my patch antenna. Their expertise and commitment played a crucial role in bringing this project to life.

\noindent Without the support and contributions of these remarkable individuals, this project would not have been possible. Thank you all for your indispensable help and for being a significant part of this accomplishment.

\clearpage
\begin{center}
\LARGE\textbf {Statement of Authorship}
\end{center}
\vspace{5pt}
I, Xunchi Ma\\
State that this work submitted for assessment is my own and expressed in my own words.Any uses made within it of works of other authors in any form (eg. Ideas, figures, text, tables) are properly acknowledged at their point of use. A list of the references employed is included.\\
Date: 18/4/2025


{
    \setstretch{1}
    \hypersetup{linkcolor=black}
    \tableofcontents
    \listoftables 
    \listoffigures 
    \glsaddall 
    \printnoidxglossary[type=\acronymtype, title=Glossary] 
}


\clearpage

\pagestyle{chapter}

\chapter{Introduction}
\chaptermark{Introduction} 

{
\hypersetup{linkcolor=black}
\minitoc
}

\section{Background \& Aim}
\noindent The rapid development of electronic devices has raised the demand for highly compact patch antennas resonating at different frequencies simultaneously. For example, the micro base station embedded in the common city infrastructure, the portable devices that can be easily placed in the human pocket. The aim of this project is to design a patch antenna which resonate in two different frequencies with low frequency at 2.25 GHz and high frequency at 5.7 GHz. To achieve the dual-band oscillation property, there are two approaches under consideration: (1) a slot applied on the primary mono-band patch antenna which creates the secondary oscillation frequency. (2) stack a second smaller patch antenna above the primary antenna. Two patch antennas work simultaneously, acquiring a dual band oscillation property. Both designs are supposed to operate with a single feed to reduce the thickness of the entire design.

\section{Research Procedure}
\subsection{Slotted Design}
\noindent In the process of advancing the project, the slot method was gradually abandoned because its poor performance compared to the stacked design. For the slot version of design, in theory, a three segments rectangular slot is carved on the original patch antenna design, which resonates at 5.8 GHz, the slot itself will create the secondary 2.4 GHz resonance frequency. However, during the simulation process, it is observed that the patch antenna operates well in lower band. But the original resonation frequency is destroyed, which does not satisfy the dual-band property. The detailed performance will be discussed later.
\subsection{Stacked Design}
\noindent After the slot method was abandoned. The focus of the research was shifted to the stacked design. At first, the overall approach was simple. The primary antenna dimension was modified to adapt to the increased thickness of the antenna. Then a secondary patch was sandwiched in the middle of the substrate and placed right below primary antenna. But the simulation result showed that the resonation pattern was not perfect. Since there is no electrical connection between to layers of patches the radiation energy is transferred purely magnetically. There are some coupling issues to be solved. To solve this problem, a methodology was introduced. It states that the vertical displacement of the align point between two patches can affect the coupling coefficient, and furthermore, affect the overall reflection coefficient of the patch antenna. So, a 5 mm displacement was introduced between two patches. And the simulation results proved that the feasibility of this methodology.

\noindent The design was then exported to be fabricated by the workshop. The actual performance of the antenna in real life still needs further measurement.

\section{Achieved Results}
\noindent After conducting the simulation for the patch antenna design. The overall performance of the antenna is summarised as follows.

\noindent For the lower bands, the S11 parameter reached -20dB, the gain reached more than 8dBi

\noindent For the upper bands, the S11 parameter reached -14dB, the gain reached more than 6dBi.

\noindent And for both resonation frequency reached the radiation efficiency around -1.3 to -1.5dB.

\noindent All the performance above showed that the prototype itself is very practical.

\chapter{Background}
\chaptermark{Background}

{
\hypersetup{linkcolor=black}
\minitoc
}
\section{Related Concepts}
\subsection{Microstrip}
\noindent A microstrip is a flat transmission line structure used in microwave and RF circuit design, particularly for applications where low cost, compactness, and ease of integration are crucial\cite{maloratsky2000reviewing}. It consists of a thin metal conductor (the signal trace), which is placed on a dielectric substrate that has a ground plane underneath. The dielectric substrate serves as the medium through which the electromagnetic fields propagate. The ground plane, typically a continuous metal layer, lies parallel to the signal trace and is typically attached to the opposite side of the substrate or embedded within it, depending on the design. The microstrip structure is widely employed in printed circuit boards (PCBs) for RF and microwave signal transmission, as well as in various communication systems such as mobile phones, satellite communication, and radar systems.

\noindent The microstrip transmission line is essentially a type of planar transmission line, where the signal trace runs on one side of a dielectric substrate, while the ground plane lies directly beneath it, separated by the thickness of the dielectric material. This construction allows for low-profile designs, and because it can be easily fabricated using conventional PCB manufacturing processes, microstrips are highly cost-effective and suitable for mass production. Microstrip lines can also be tailored for different applications by adjusting parameters such as the width of the conductor, the dielectric constant of the substrate, and the thickness of the substrate.

\noindent In terms of electromagnetic theory, microstrip transmission lines are considered a form of waveguide, though the structure is significantly more compact. The signal on a microstrip behaves similarly to a guided wave, where electromagnetic energy is confined to the region between the conductor and the ground plane. The field distribution in a microstrip is non-uniform; the electric field is primarily confined near the surface of the conductor and the dielectric material, while the magnetic field is distributed around the conductor. As a result, microstrip lines exhibit both transverse electric (TE) and transverse magnetic (TM) field components, with the dominant mode being a quasi-TEM (transverse electromagnetic) mode, which allows for efficient signal transmission with minimal losses at microwave frequencies.

\noindent The characteristic impedance of a microstrip is a crucial parameter, as it governs the impedance matching between the microstrip and other circuit elements, such as antennas, amplifiers, or other transmission lines. This impedance is dependent on several factors, including the width of the microstrip trace, the dielectric constant of the substrate, and the thickness of the substrate. A commonly targeted characteristic impedance is 50 ohms, which is standard for many RF and microwave systems. The relationship between these physical dimensions and the characteristic impedance can be calculated or determined using empirical formulas or numerical simulations.

\noindent One of the primary advantages of microstrip transmission lines is their simplicity in design and integration. They can be easily incorporated into the same PCB as other circuit elements, such as active devices, passive components, or connectors, without the need for complex interconnects. This integration leads to reduced manufacturing complexity and cost, as well as smaller form factors. Additionally, microstrip lines are lightweight and flexible, making them suitable for compact, portable systems where space and weight are at a premium.

\noindent However, microstrips also have certain limitations that must be addressed in high-frequency applications. One of the major challenges is the effect of losses, particularly in the dielectric substrate and the conductor itself. At high frequencies, skin effect causes current to concentrate on the surface of the conductor, increasing resistance and leading to higher losses. Additionally, dielectric losses are a concern, especially when using substrates with lower-quality dielectric materials. These losses can degrade the signal quality and reduce the efficiency of the transmission line. Another issue is the potential for radiation losses, which can arise if the microstrip is not properly shielded or if the signal is being transmitted over a longer distance.

\noindent The performance of a microstrip transmission line is also affected by the properties of the dielectric substrate. The dielectric constant (also referred to as the relative permittivity) of the material influences the velocity of the wave propagation and, consequently, the physical dimensions of the microstrip required to achieve a specific characteristic impedance. High dielectric constant materials reduce the required width of the microstrip trace but may increase losses, whereas low dielectric constant materials allow for a larger trace width but are often associated with lower loss and better performance at higher frequencies. The substrate thickness also affects the microstrip's impedance and the distribution of the electric and magnetic fields. Thin substrates lead to a more confined field distribution, while thicker substrates spread the fields more, impacting the efficiency of the signal transmission.

\noindent To mitigate these challenges, advanced microstrip designs may involve the use of multilayer PCB structures, where the microstrip is placed on different layers, providing more control over the signal propagation and enabling more complex circuits. Alternatively, the use of low-loss dielectric materials or surface treatments on the conductors can reduce the impact of losses and enhance the overall performance of the transmission line. Another approach is to employ shielded microstrips or techniques like coaxial feeding, where the signal is better contained and isolated from external interference, which can be especially important in high-sensitivity applications like radar and satellite communication.

\noindent Microstrip lines also enable various specialized applications, such as the development of components like couplers, filters, power dividers, and baluns. These components exploit the unique characteristics of microstrip transmission lines to provide controlled signal splitting, filtering, and impedance matching. For example, a Wilkinson power divider is a passive device often designed using microstrip technology that divides an input signal into multiple outputs with minimal loss and good impedance matching, which is widely used in RF and microwave systems.

\noindent So, microstrip transmission lines form the backbone of many RF and microwave circuit designs due to their simplicity, cost-effectiveness, and ease of integration into modern electronic systems. Their ability to deliver high-performance signal transmission in compact, low-profile packages makes them indispensable in fields ranging from telecommunications and aerospace to consumer electronics. However, achieving optimal performance requires careful consideration of factors such as substrate material, thickness, loss characteristics, and impedance matching, especially at high frequencies. Despite these challenges, ongoing advancements in material science and fabrication techniques continue to improve microstrip technology, expanding its range of applications and enabling more efficient, reliable communication systems.

\subsection{Patch antenna}
\noindent A patch antenna, often referred to as a microstrip antenna, is a type of low-profile radiator widely used in modern wireless communication systems due to its compact form factor, ease of fabrication, and relatively low cost\cite{lee2012microstrip}. At its most basic level, a patch antenna consists of a thin conductive patch—commonly made of copper or another suitable metal—placed on top of a dielectric substrate that is backed by a ground plane. The patch itself can assume a variety of shapes—rectangular, square, circular, elliptical, triangular, or even more complex geometries—but rectangular and circular patches are among the most popular for straightforward design and analytical treatment. The dielectric substrate plays a critical role in determining the antenna’s electrical characteristics, including resonant frequency, bandwidth, and radiation efficiency, owing to its thickness and relative permittivity. A thinner substrate with higher permittivity typically leads to a more compact design but reduces bandwidth and radiation efficiency, whereas a thicker substrate with lower permittivity can enhance bandwidth at the expense of an increased overall size and possible unwanted surface-wave effects.

\noindent From an operational standpoint, patch antennas rely on the principle of resonant behavior. When an RF signal is fed to the patch—usually by a microstrip line, coaxial probe, proximity coupling, or aperture coupling—the patch forms a resonant cavity with the ground plane. The fringing fields at the edges of the patch are responsible for the antenna’s radiation; effectively, the patch and the ground plane function like an open-ended resonant transmission line, where the standing waves generate electromagnetic fields that “leak” or radiate into free space. This operation is fairly narrowband under typical single-layer, single-resonance configurations. However, designers can modify various aspects—patch dimensions, feed placement, or geometry—to achieve a particular resonant frequency and a desired input impedance match.

\noindent One of the challenges with patch antennas is their relatively narrow bandwidth, often just a few percent of the center frequency, making them inherently less versatile when broad or multiband coverage is required. Over the years, numerous techniques have been developed to counter this limitation. For instance, the use of thicker substrates, air-gap layers, or stacked patches significantly increases the operational bandwidth by introducing additional resonances or by broadening the original resonance. Another design strategy employs slots in the patch to create multiple current paths, thereby allowing the antenna to resonate at multiple frequencies or achieve broader bandwidth. Additionally, employing different feed mechanisms—such as aperture coupling or electromagnetic coupling—can provide further degrees of freedom to optimize impedance matching and improve overall efficiency across a wider band.

\noindent Radiation characteristics are also critical in patch antenna design. A conventional patch antenna usually exhibits a broadside radiation pattern, where the main lobe is oriented perpendicular to the plane of the patch, and the gain can vary depending on substrate thickness, losses, and operating frequency. Because patch antennas are inherently planar, they lend themselves to easy integration with integrated circuits and phased arrays. In phased array configurations, many patches can be arranged in a grid, allowing electronic beam steering or shaped-beam capabilities that cater to various wireless, radar, and satellite communication needs. Antenna engineers pay close attention to gain, return loss, front-to-back ratio, side-lobe levels, and polarization purity when tailoring a patch antenna or array to specific performance criteria.

\noindent Moreover, the mechanical robustness, low manufacturing cost, and compatibility with printed circuit board (PCB) technology make patch antennas indispensable in handheld devices, embedded systems, and vehicular or aerospace applications. Designers often incorporate conformal or flexible substrates that enable patch antennas to attach to curved surfaces or be embedded within non-planar platforms. Despite their advantages, patch antennas can experience drawbacks related to low power-handling capability and potential losses within the substrate material, which must be accounted for in high-power or high-frequency scenarios (e.g., millimeter-wave systems). Nonetheless, ongoing research continues to refine novel geometries, feeding schemes, and material choices that expand the performance envelope of patch antennas, keeping them at the forefront of contemporary microwave and millimeter-wave engineering.

\subsection{Multi-band Patch Antenna}
\noindent A multi-band patch antenna is a specialized type of microstrip antenna that is designed to operate efficiently across two or more frequency bands\cite{jhamb2011novel}. The primary goal of multi-band patch antennas is to enable communication systems to operate in multiple frequency bands without requiring multiple antennas or complex switching mechanisms. These antennas are widely used in modern wireless communication systems, including mobile phones, GPS, Wi-Fi, Bluetooth, and satellite communications, as they allow for the simultaneous operation of various communication protocols within different frequency ranges. The unique challenge in the design of multi-band patch antennas lies in achieving good performance across these multiple bands while maintaining compactness, efficiency, and low cost.

\noindent The special characteristic of multi-band patch antennas is that they can resonate at more than one frequency, which is essential for meeting the diverse needs of modern communication systems. Unlike traditional single-band patch antennas, which are designed to resonate at a single frequency, multi-band patch antennas leverage various techniques to create additional resonant modes at different frequencies. These techniques can include the use of additional radiating elements, modifications to the patch geometry, multiple layers (stacked patches), and the introduction of slots or parasitic elements. By carefully controlling these factors, designers can create antennas that exhibit multiple resonant frequencies while minimizing interference between the different bands.

\noindent One of the most common techniques used in the design of multi-band patch antennas is the introduction of slots within the patch. These slots act as radiating elements that modify the current distribution on the patch and enable the antenna to resonate at different frequencies. The dimensions and positioning of these slots can be optimized to create specific resonances in the desired frequency bands. In some cases, the slots can be placed in strategic locations to create resonances in the lower or higher frequency ranges, allowing the antenna to support communication standards like GSM, UMTS, LTE, Wi-Fi, and Bluetooth simultaneously.

\noindent Another widely employed method is the use of stacked patches, where two or more patch elements are placed on top of each other with different dielectric layers in between. The different resonant frequencies are achieved by controlling the dimensions and separation of the patches. Stacked patch antennas have the advantage of compactness, as the use of multiple layers enables the antenna to achieve multiple frequency bands without occupying excessive space. The electromagnetic coupling between the stacked patches allows for the generation of different resonant modes, each corresponding to a distinct frequency band. Additionally, the use of stacked configurations can improve the overall gain and bandwidth of the antenna.

\noindent The geometry of the patch itself can also be modified to support multi-band operation. For example, the traditional rectangular or square-shaped patches can be altered to include shapes like triangular, circular, or fractal patches, which have multiple resonant modes. Fractal shapes, in particular, have gained attention in the design of multi-band antennas because of their self-similar nature, which allows them to resonate at multiple frequencies within a compact area. By adjusting the dimensions of the fractal elements, the antenna can be tuned to operate at different frequency bands while maintaining a small footprint.

\noindent Multi-band patch antennas can also take advantage of the concept of parasitic elements, which are passive components that are placed near the radiating element but are not directly connected to the feed. These parasitic elements, such as directors or reflectors, can be arranged around the main patch to modify the radiation pattern and enhance the antenna's performance at specific frequencies. The placement and size of the parasitic elements can be optimized to introduce additional resonant frequencies, making the antenna suitable for multi-band operation.

\noindent Impedance matching is another critical aspect of multi-band patch antenna design. In a multi-band configuration, the antenna must be designed to match the impedance of the transmitter or receiver at each of the different resonant frequencies. This can be achieved by adjusting the feed mechanism or by using impedance-matching networks. A well-matched impedance ensures maximum power transfer and minimizes signal reflection at each operating frequency. In practice, the feed mechanism is often adapted to ensure that the antenna can operate efficiently across all desired frequency bands, sometimes employing techniques like multi-port feeding, where different feed points are used for different frequency bands.

\noindent The use of multi-band patch antennas offers several key advantages. First, they significantly reduce the need for multiple antennas, which is particularly beneficial in applications where space is limited, such as in handheld devices and mobile platforms. Additionally, multi-band antennas allow for more efficient use of the available spectrum, enabling devices to support multiple wireless standards and frequency bands without the need for separate antenna elements for each band. This is especially important in the context of modern smartphones and IoT devices, which often need to support a wide variety of communication protocols, including cellular networks, Wi-Fi, Bluetooth, and GPS, among others.

\noindent Despite these advantages, there are several challenges associated with the design of multi-band patch antennas. One of the main challenges is ensuring good isolation between the different frequency bands. Since the antenna operates in multiple frequency ranges, there is a risk that the resonant modes of different bands could interfere with each other, leading to poor performance. This can be mitigated by careful optimization of the antenna geometry, feed network, and substrate materials, ensuring that each resonant mode is sufficiently separated and does not negatively impact the other bands.

\noindent Another challenge is the trade-off between antenna size and performance. While multi-band patch antennas allow for compact designs, there is often a compromise between achieving wide bandwidth and maintaining a small form factor. Achieving a wide bandwidth across multiple frequency bands can require larger patch dimensions or more complex geometries, which may increase the overall size of the antenna. Designers must carefully balance these factors to meet the size and performance requirements of the intended application.

\noindent In conclusion, multi-band patch antennas are a powerful solution for modern communication systems that require support for multiple frequency bands within a compact, efficient, and cost-effective design. By leveraging techniques such as slotting, stacking, geometry modifications, and parasitic elements, multi-band patch antennas can meet the demands of a variety of wireless applications, including mobile communications, satellite systems, and IoT. However, careful attention to impedance matching, isolation between bands, and optimization of the antenna geometry is necessary to ensure that the antenna performs well across all operating frequencies. As communication technologies continue to evolve, multi-band patch antennas will remain a key component in the development of next-generation wireless devices and systems.
\section{Previous Works}

\noindent Recent research on microstrip patch antennas has focused on enhancing bandwidth, gain, polarization diversity, and integration with wearable and compact platforms. In \cite{RN5}, Fernandes et al. present a dual‐band patch antenna operating at 2.4 and 5.8 GHz that incorporates a Frequency Selective Surface (FSS) reflector to improve radiation characteristics. Their design employs a truncated ground plane and a novel integration of a single‐layer FSS screen—designed with a double square loop geometry—to boost gain, directivity, and front‐to‐back ratio. Simulation and experimental measurements confirm improvements of up to 5.2 dB in gain at 2.4 GHz, demonstrating the effectiveness of FSS integration in dual‐band applications.

\noindent In a different approach, Abdullahi and Obagidi in \cite{srivastava2015design} focus on the design and parametric analysis of a rectangular microstrip patch antenna over multiple frequency bands ranging from L to Ku. Utilizing the transmission line model, they derive closed‐form expressions to calculate key antenna parameters—including patch dimensions and substrate height—and show that both the antenna size and bandwidth are highly dependent on the operating frequency and dielectric constant. Their study provides practical guidelines for antenna designers by revealing that a lower dielectric constant can yield wider bandwidth and enhanced efficiency, a critical insight for multi‐band wireless applications.

\noindent Addressing the demands of in‐band full‐duplex (IBFD) systems, Bui et al. in \cite{RN4} propose a dual‐band dual‐polarized slotted-patch antenna with a low-profile structure. Their design features a 45°-rotated square patch with capacitive probes, which, through a double differential feed, excites the TM10 mode for the lower resonance and employs four trapezoidal slots to generate a second resonance at a higher frequency. The design is further enhanced by a dual-band rat-race coupler, achieving high port-to-port isolation (exceeding 40 dB) and measured gains up to 10.7 dBi. This work underscores the feasibility of integrating dual-polarized and IBFD capabilities within a compact patch antenna architecture.

\noindent For wireless body area network (WBAN) applications, Hong et al. in \cite{hong2014dual} introduce a dual‐band, dual‐mode patch antenna designed specifically for on–on–off communication scenarios. By employing a half-circular patch with shorting posts to excite a higher-order TM41 mode and incorporating a ground slot for an additional resonance, the antenna achieves dual-mode operation. The half-mode technique effectively reduces the overall antenna size while providing a vertical monopole-like pattern at 2.45 GHz (suitable for on-body communication) and a horizontal pattern at 5.8 GHz (for off-body links). Experimental validation on muscle-equivalent phantoms further demonstrates its practical viability in WBAN environments.

\noindent Expanding on wearable technology, Zhou et al. in \cite{RN6} propose a dual‐band and dual‐polarized circular patch textile antenna tailored for on-/off‐body WBAN applications. Fabricated on a flexible felt substrate, the antenna leverages a single circular patch with eight strategically placed slots to simultaneously achieve circular polarization at 2.45 GHz (enhancing link reliability for off-body communications) and vertical polarization at 5.8 GHz (for on-body interactions). The design’s low profile, combined with its resilience under bending and compliance with SAR standards, highlights its potential for seamless integration into wearable devices.

\noindent Collectively, these works demonstrate a broad spectrum of strategies—from FSS integration and transmission line-based design to advanced dual-polarized feeding schemes and textile implementations—addressing the multifaceted challenges of modern antenna design. They not only advance the state of the art in performance enhancement and multi-band operation but also provide valuable insights into the trade-offs between antenna size, polarization, and bandwidth in both conventional and wearable applications, proving that it is feasible for the fabrication of dual band patch antenna for this project.



\chapter{My Work}
\chaptermark{My Work}

{
\hypersetup{linkcolor=black}
\minitoc
}

\section{Initiation}

\noindent In the initial phase of the project, the aim of the project was determined, which is fabricating a dual band patch antenna. For the compliance of the patch antenna, both bands were set in ISM radio band. It is the radio spectrum reserved internationally for industrial, scientific, and medical (ISM) purposes, excluding applications in telecommunications. In this way, the unnecessary disruption and electromagnetic interference could be greatly avoided.

\noindent After the resonant frequency, or the operation frequency was determined, the background research was conducted to verify the feasibility of this research approach in advance. As what is already mentioned above.

\section{Mono Band Patch Antenna}
\noindent The project entered the practical stage after the thorough planning and background research. In the first step, A simulation on mono band patch antenna was considered as necessary to verify the overall performance of a patch antenna and the detailed influence of difference parameters of the antenna.
\subsection{General Theory}
\noindent As stated in the online article \cite{3G-Aerial_Patch_Antenna} based on the article \cite{munson1988conformal},Theoretically, the dimension of the patch is determined by the half wavelength of the resonant frequency plus some offset with respect to the relative dielectric constant \(\varepsilon_R\) as well as the effective dielectric constant \(\varepsilon_{\text{eff}}\)

\[
\text{Dimension of patch} \approx \frac{\lambda}{2} + \Delta(\varepsilon_R, \varepsilon_{\text{eff}})
\]

\noindent where \(\lambda\) is the wavelength corresponding to the resonant frequency, and \(\Delta(\varepsilon_R, \varepsilon_{\text{eff}})\) is the offset dependent on the relative and effective dielectric constants. for more detailed calculation the formula is given as the following equation \ref{eq:width}, \ref{eq:eps_eff} and \ref{eq:length}.

\begin{equation}
    \text{Width} = \frac{c}{2 f_0 \sqrt{\frac{\varepsilon_R + 1}{2}}}
    \label{eq:width}
\end{equation}

\begin{equation}
    \varepsilon_{\mathrm{eff}} 
    = \frac{\varepsilon_R + 1}{2} 
    \;+\; \frac{\varepsilon_R - 1}{2} 
      \left(\frac{1}{\sqrt{\,1 + 12\,\frac{h}{W}}}\right)
    \label{eq:eps_eff}
\end{equation}

\begin{equation}
    \text{Length} 
    = \frac{c}{2 f_0 \sqrt{\varepsilon_{\mathrm{eff}}}}
    \;-\; 0.824\,h \,
      \left(\frac{\varepsilon_{\mathrm{eff}} + 0.3}{\varepsilon_{\mathrm{eff}} - 0.258}\right)
      \left(\frac{\frac{W}{h} + 0.264}{\frac{W}{h} + 0.8}\right)
    \label{eq:length}
\end{equation}
\subsection{Parameters Calculation}
\noindent By utilizing the online calculator\cite{3G-Aerial_Patch_Antenna_calculator} provided by the source of the online article\cite{3G-Aerial_Patch_Antenna}. The detailed parameters for two patch antennas operating at two desired frequencies are calculated as follows,listed in the table \ref{tab:mono_band_5.8} and \ref{tab:mono_band_2.4}.

\noindent The first two mono band antennas were designed and simulated to operate within the \gls{ism} bands, ensuring compliance with regulatory requirements for unlicensed wireless communication systems.
\subsection{Realisation}
\noindent Both designs were simulated and their detailed performance was tested and measured using the software: \gls{cst}. The soft stated that two antennas had minimum S11 parameters below -10 dB and proper bandwidth near their operation frequency. The detailed performance plots will be attached in the Appendices in order to simplify the length of the report.

\begin{figure}[H] 
    \centering 
    \includegraphics[]{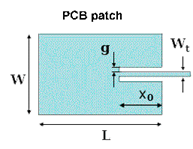} 
    \rule{35em}{0.5pt} 
    \caption {The sketch of a patch antenna on a PCB}
    \label{fig:pcb_patch} 
\end{figure}

\noindent It is worth noting from the table that there are some carves near the feed lines. As what is shown on the figure \ref{fig:pcb_patch}. The $W$ and $L$ are the width and length of the patch antenna. The $W_t$ is the width of the feed line. The $g$ and $x_0$ are the width and the length of the carved slot on the patch. Those slots improved the bandwidth\cite{wong2015bandwidth} of the patch antenna as well as matched the impedance\cite{sharma2017impedance} without increasing the physical size of the patch antenna itself.

\begin{table}[htbp]
\centering
\caption{The Dimension of the 5.8 GHz Mono Band Patch Antenna}
\label{table:patch_antenna_dimensions}
\begin{tabular}{c l c}
\toprule
\textbf{Symbol} & \textbf{Description} & \textbf{Value} \\
\midrule
$f$     & Mean frequency of the range                     & 5.8 GHz \\
$Z_o$   & Antenna input impedance                         & 50 $\Omega$ \\
$E_r$   & Dielectric constant of the substrate            & 3.55 \\
$H$     & Substrate height                                & 1.5 mm \\
$\lambda$ & Wavelength                                     & 51.7 mm \\
$PW$    & Patch width                                     & 17.1 mm \\
$PL$    & Patch length                                    & 13.1 mm \\
$Z_p$   & Patch input impedance $Z_p$                     & 182 $\Omega$ \\
$X_0$   & Distance to match input impedance 50 $\Omega$   & 4.3 mm \\
$G$     & Gap width                                       & 0.38 mm \\
$Wt$    & Microstrip ‘feeder’ width                       & 3 mm \\
\bottomrule
\end{tabular}
\label{tab:mono_band_5.8}
\end{table}

\begin{table}[htbp]
\centering
\caption{The Dimension of the 2.4 GHz Mono Band Patch Antenna}
\label{table:bottom_patch_antenna_dimensions}
\begin{tabular}{c l c}
\toprule
\textbf{Symbol} & \textbf{Description} & \textbf{Value} \\
\midrule
$f$     & Mean frequency of the range                     & 2.4 GHz \\
$Z_o$   & Antenna input impedance                         & 50 $\Omega$ \\
$E_r$   & Dielectric constant of the substrate            & 3.55 \\
$H$     & Substrate height                                & 1.5 mm \\
$\lambda$ & Wavelength                                     & 125 mm \\
$PW$    & Patch width                                     & 41.4 mm \\
$PL$    & Patch length                                    & 32.7 mm \\
$Z_p$   & Patch input impedance $Z_p$                     & 181 $\Omega$ \\
$X_0$   & Distance to match input impedance 50 $\Omega$   & 10.6 mm \\
$G$     & Gap width                                       & 0.86 mm \\
$Wt$    & Microstrip ‘feeder’ width                       & 2.9 mm \\
\bottomrule
\end{tabular}
\label{tab:mono_band_2.4}
\end{table}

\section{Slotted Dual-band Patch Antenna Design}
\subsection{Feasibility Analysis}
\noindent After the feasibility of the mono band patch antenna was successfully verified, the next consideration of the project was to add another resonant frequency for the mono band patch antenna to help it acquire the dual band property.
\noindent The first approach to be considered is the slotted method. The introduction of a slot in a microstrip patch antenna effectively modifies the current distribution and alters the antenna’s resonant modes, allowing it to operate effectively at more than one frequency band. Specifically, the slot acts as a perturbation in the patch that creates an additional electrical path, leading to a second, distinct resonance in conjunction with the original patch resonance. This phenomenon can be understood by examining the surface current behavior: the slot forces a redistribution of currents on the radiating surface, effectively exciting another resonant mode at a different frequency than the primary patch mode. By appropriately tuning the slot geometry (e.g., length, width, and position), designers can optimize the two resonant frequencies to fall within designated bands, resulting in a dual-band or even a multi-band antenna. This approach to achieving multi-band performance is favored for its compactness and simplicity, requiring relatively minimal changes in fabrication compared to more complex multilayer or stacked configurations.
\subsection{Design Procedure}
\begin{figure}[H] 
    \centering 
    \includegraphics[width=1.0\textwidth]{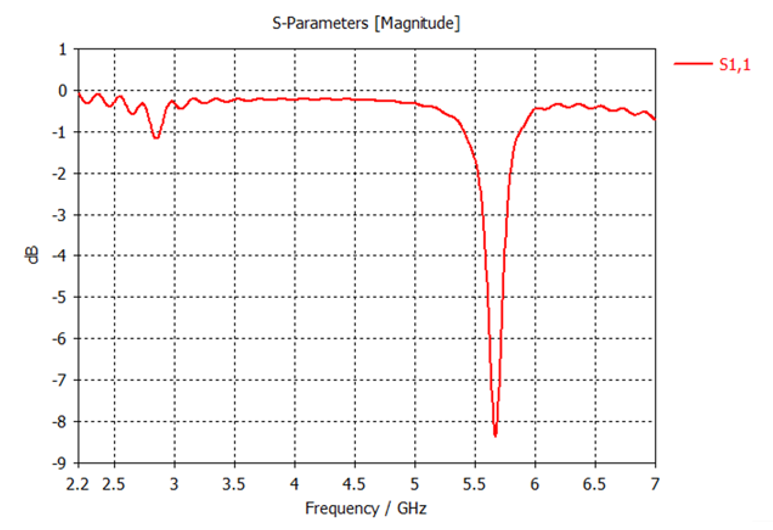} 
    \rule{35em}{0.5pt} 
    \caption {The S11 simulation result for slotted design}
    \label{fig:slotted_s11} 
\end{figure}

\begin{figure}[H] 
    \centering 
    \includegraphics[width=1.0\textwidth]{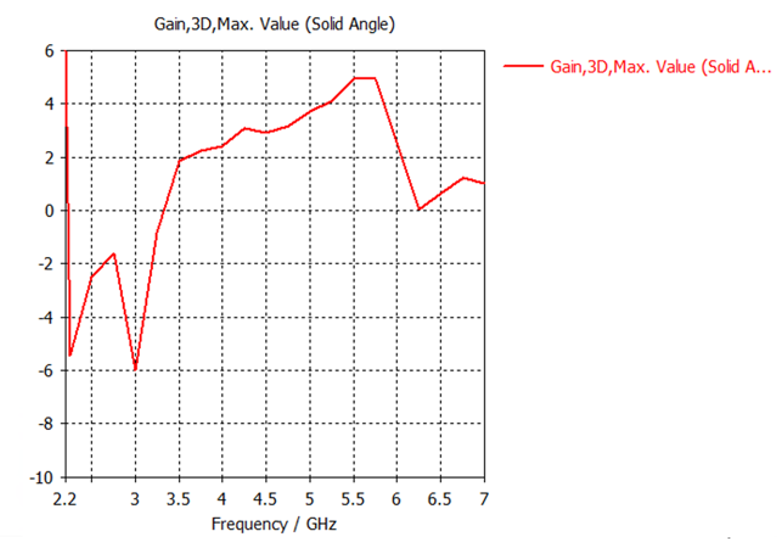} 
    \rule{35em}{0.5pt} 
    \caption {The gain simulation result for slotted design}
    \label{fig:slotted_gain} 
\end{figure}

\begin{figure}[H] 
    \centering 
    \includegraphics[width=1.0\textwidth]{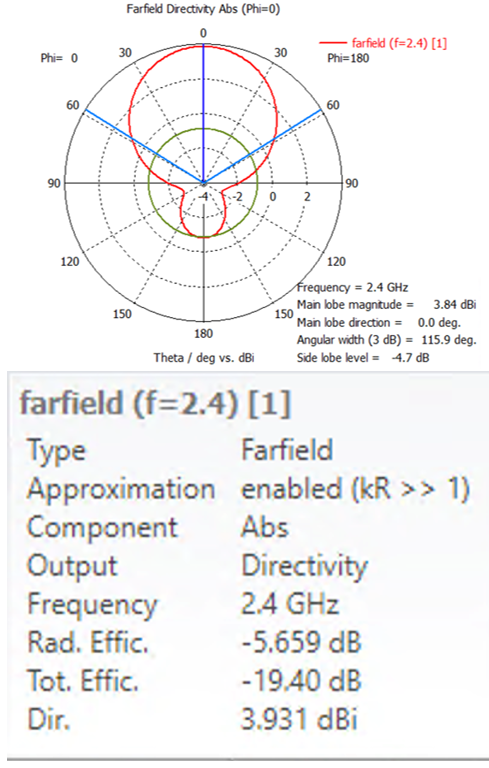} 
    \rule{35em}{0.5pt} 
    \caption {The lower band farfield simulation result for slotted design}
    \label{fig:slotted_2.4_farfield} 
\end{figure}

\begin{figure}[H] 
    \centering 
    \includegraphics[width=1.0\textwidth]{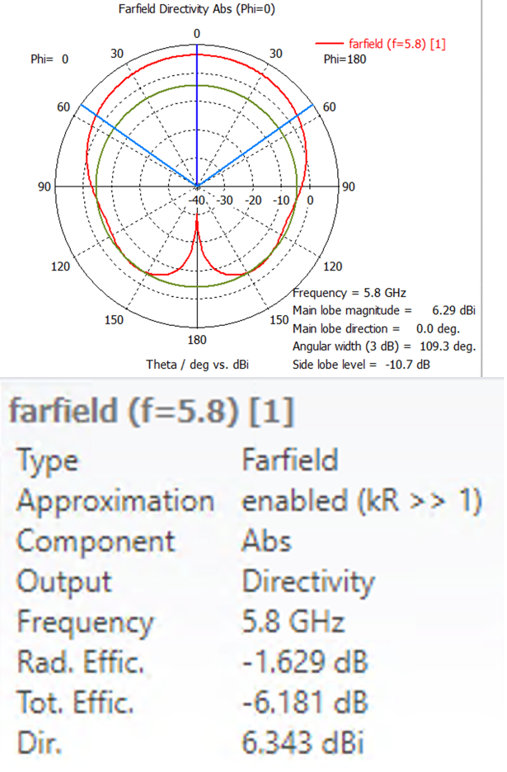} 
    \rule{35em}{0.5pt} 
    \caption {The upper band farfield simulation result for slotted design}
    \label{fig:slotted_5.8_farfield} 
\end{figure}

\noindent On the base patch antenna whose resonant frequency is 5.8 GHz and the detailed parameters is as same as the table \ref{tab:mono_band_5.8}, a slot was carved to create a second resonate frequency. The total length of the slot is approximately the quarter of the wavelength of 2.4 GHz. After thoroughly adjustment, 35 mm was chosen to be the total length of the slot and the 1 mm to be the width. But the length of the slot exceeds both the width and length of the base patch. To solve this problem, a U-shape or a two-bend structure was introduced to fit the slot into the patch.
\noindent After modeling in the CST software, the dimension of the patch is shown in the figure \ref{fig:slotted_design} below. The unit of figure is 0.1 mm.
\begin{figure}[H] 
    \centering 
    \includegraphics[width=0.7\textwidth]{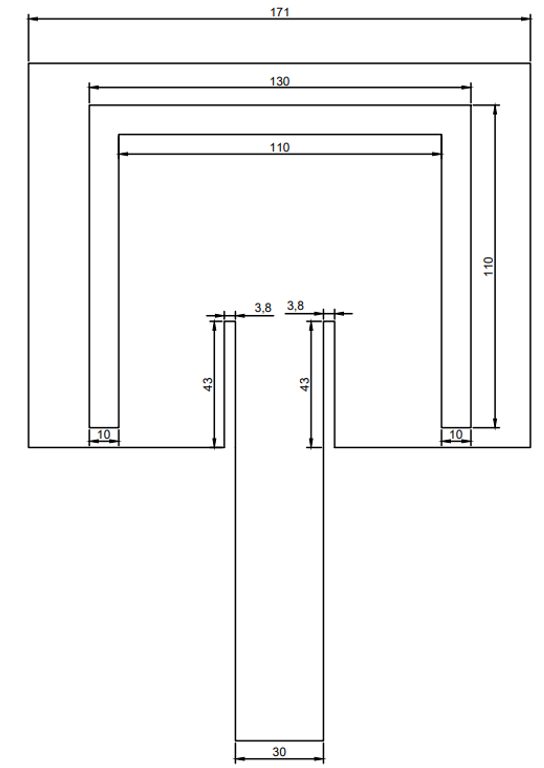} 
    \rule{35em}{0.5pt} 
    \caption {The slotted patch antenna dimension}
    \label{fig:slotted_design} 
\end{figure} 
\noindent After conducted the simulation of the slotted patch antenna in the CST software. the S11 simulation result is shown in the figure \ref{fig:slotted_s11}, the gain result is shown in the figure \ref{fig:slotted_gain}, farfield in \ref{fig:slotted_2.4_farfield} and \ref{fig:slotted_5.8_farfield}. 

\noindent As it can be observed that, for S11 parameter, in the lower band, nearly all of the signal was reflected, with a minimum S11 at around -1dB, and for the upper band, the original resonant pattern was also destroyed, the minimum S11 parameter could not reach -10dB, meaning the patch antenna itself does not have a effective bandwidth for the operation. For the gain, at lower band, the patch antenna even had a negative gain, which is totally not usable. For the farfield, the antenna did have some directivity. But at lower band the efficiency of the antenna was too low, making it unusable.

\noindent Since that was the already the optimized result. The slotted design approach was abandoned during the promotion of the project.

\section{Stacked Dual-band Patch Antenna Design}
\subsection{Feasibility Analysis}
\noindent After the feasibility of the slotted patch antenna for this project was denied in the previous work, a new approach was proposed to reach the dual band property, which is the stacked structure.

\noindent A stacked patch configuration enables dual-band operation by layering multiple resonant elements in a compact structure, each of which can be designed to support a specific frequency of interest. In this configuration, the electromagnetic coupling between the stacked patches plays a critical role in exciting distinct resonance modes. Proper selection of patch dimensions, dielectric layers, and inter-layer separations allows designers to fine-tune these modes, ensuring minimal interference while enhancing the overall bandwidth and gain at the two targeted frequency bands. Furthermore, the stacked configuration can mitigate the limitations associated with single-layer antennas, such as narrow bandwidth and limited design flexibility, by exploiting the multi-layer arrangement to spread and manage surface currents. As a result, a stacked patch antenna can operate efficiently at two separate resonant frequencies, which is feasible for the aim of this project.
\subsection{Initial Design Procedure}
\begin{figure}[H] 
    \centering 
    \includegraphics[width=1.0\textwidth]{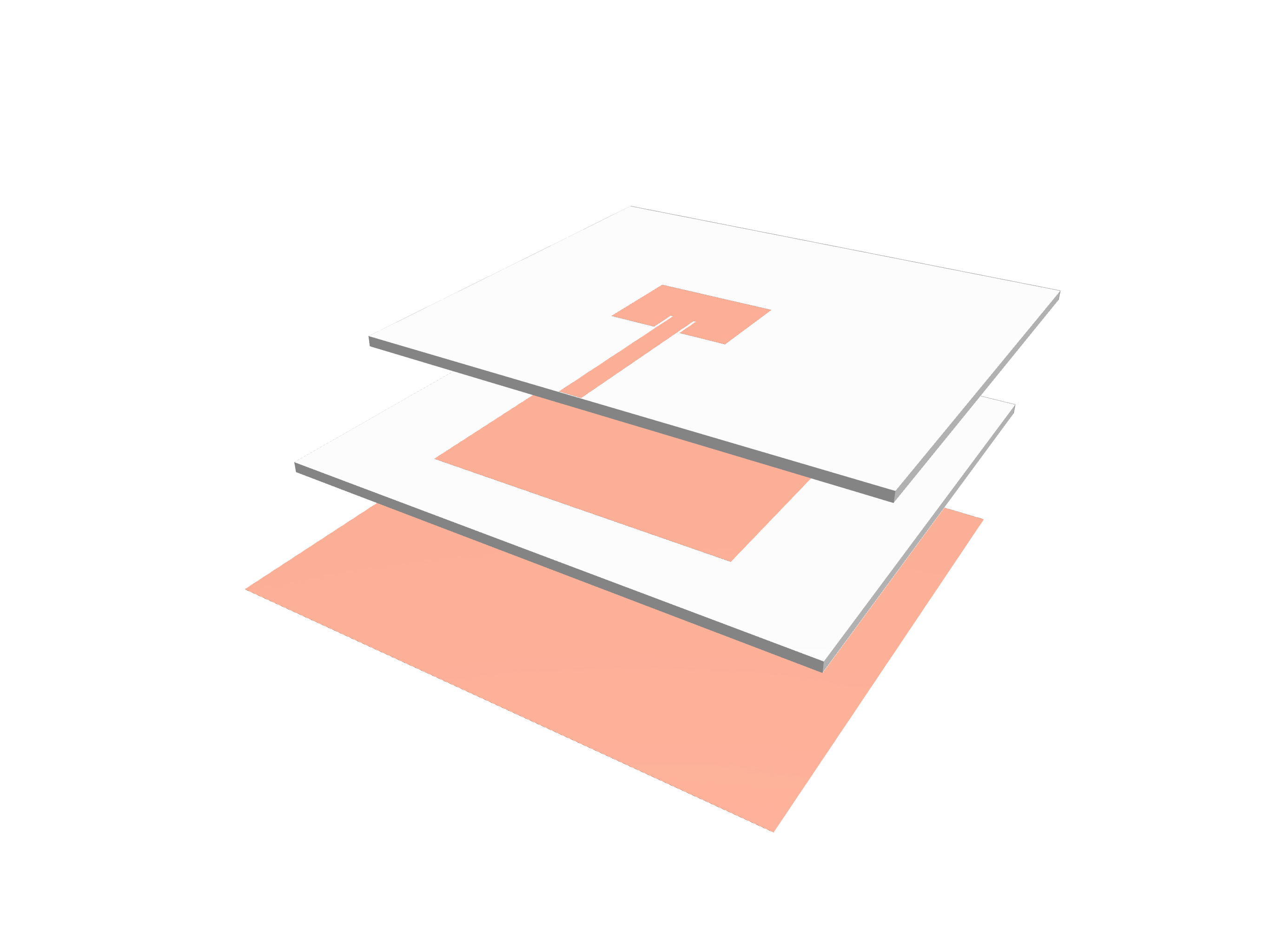} 
    \rule{35em}{0.5pt} 
    \caption {The 3D rendered image for stacked patch antenna}
    \label{fig:stack_3d} 
\end{figure}

\noindent The initial methodology and approach for the stacked dual band patch antenna was simple. As shown in the figure \ref{fig:stack_3d},this dual-band patch antenna was designed to operate in the 2.25–2.35 GHz and 5.6–5.8 GHz frequency bands by employing a stacked configuration with two layers of substrate and three layers of copper, forming a copper–substrate–copper–substrate–copper structure. A single feed, connected to the side rather than the bottom, excites the top patch to achieve resonance at 5.7 GHz and simultaneously helps minimize the antenna’s overall thickness. The larger middle patch was responsible for generating the lower 2.3 GHz band, while the bottom copper layer serves as the ground plane. There was no electrical connection between different layers of copper. The small patch antenna on the top layer was coupling purely magnetically with the large patch sandwiched in the middle layer.

\begin{figure}[H] 
    \centering 
    \includegraphics[width=0.8\textwidth]{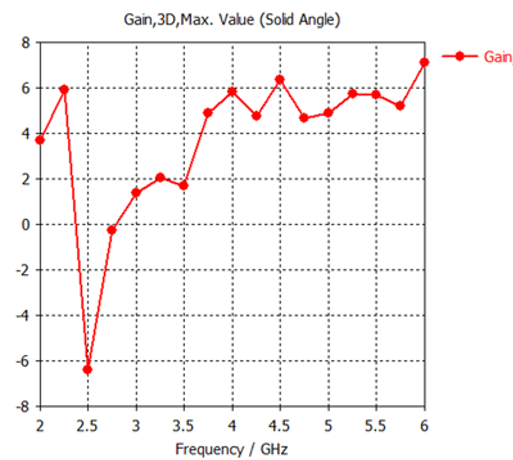} 
    \rule{35em}{0.5pt} 
    \caption {The gain simulation result for the initial stacked design}
    \label{fig:stack_gain} 
\end{figure}

\begin{figure}[H] 
    \centering 
    \includegraphics[width=0.8\textwidth]{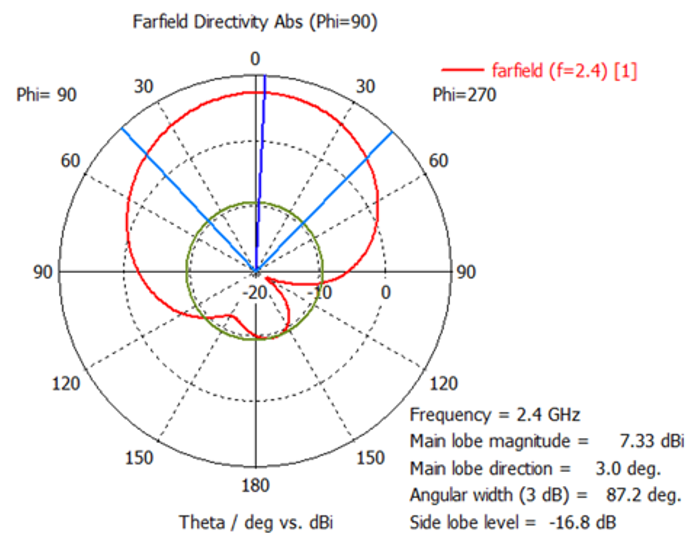} 
    \rule{35em}{0.5pt} 
    \caption {The lower band farfield simulation result for the initial stacked design}
    \label{fig:stack_farfield_2.4} 
\end{figure}

\begin{figure}[H] 
    \centering 
    \includegraphics[width=0.8\textwidth]{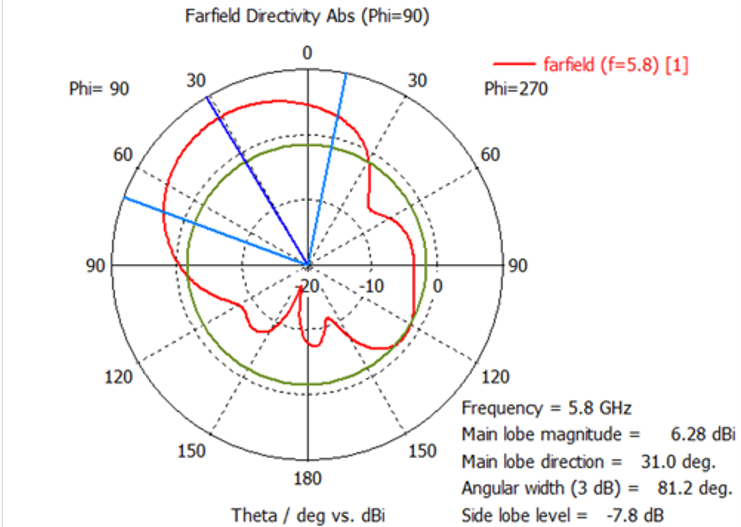} 
    \rule{35em}{0.5pt} 
    \caption {The upper band farfield simulation result for the initial stacked design}
    \label{fig:stack_farfield_5.8} 
\end{figure}

\noindent After this initial design was modeled and simulated in the CST software and the dimension of the two layers of patches were slightly adjusted according to the parameters listed in the table \ref{tab:mono_band_5.8} and \ref{tab:mono_band_2.4} in order to further fine tuning the performance. The output simulation result is shown in the figures below. Figure \ref{fig:stack_s11} for the S11 parameter, figure \ref{fig:stack_gain} for the gain, figure \ref{fig:stack_farfield_2.4} and \ref{fig:stack_farfield_5.8} for the farfields.

\begin{figure}[H] 
    \centering 
    \includegraphics[width=0.8\textwidth]{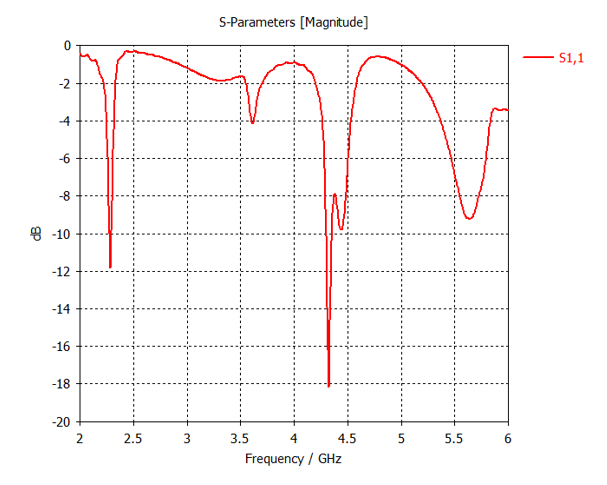} 
    \rule{35em}{0.5pt} 
    \caption {The S11 simulation result for the initial stacked design}
    \label{fig:stack_s11} 
\end{figure} 
\subsection{Initial Design Analysis}
\noindent As it can be observed from the simulation from the previous section that, the stacked dual-band patch antenna did resonate at multiple different frequencies. But the upper band was not as desired.The upper band is around 4.3 GHz, which was unacceptable. Also for the lower band, the resonant frequency was very close to the goal of this project. But the effective bandwidth, which is the S11 section below -10 dB, was not a feasible during the actual operation of the antenna. The antenna will easily jump out of operating range if there is some jitter for the carrier frequency.

\noindent For the Gain of the patch antenna, it was good at exact the resonant frequency. But right near the lower band, there was even some negative gain. Regarding that the directivity for this patch antenna could not be negative at the positive perpendicular direction. The radiation efficiency for this frequency might be very low. Or the matching for the feed line was very bad.

\noindent When it comes to the farfield for this antenna at phi=90, the lower band looked fairly good. But for the upper band, it can be observed that the main lobe of the antenna pointed all the way upwards. For the antenna, the main lobe should have aligned withe normal line of the substrate, or in other words, perpendicular to the patch plain in order to better point to the transceiver side to reach high SNR and transmission speed.
\subsection{Correction for the Initial Design}
\subsubsection{Methodologies}
\noindent In order to further correct the defects mentioned above for this stacked dual band patch antenna, new methodologies were adopted. There was an article\cite{bulja2023microstrip} states that  the displacement of different layers in a microstrip antenna—particularly the relative alignment between the patch, substrate, and ground plane—plays a significant role in affecting electromagnetic coupling, which directly influences antenna performance. When layers are not perfectly aligned or become slightly displaced, unintended coupling may occur between the patch and the feedline or between adjacent resonant structures. This misalignment alters the distribution of the electromagnetic fields and can lead to impedance mismatches, increased losses, and degraded radiation patterns. For instance, if the feedline is not properly centered with respect to the patch, it may cause asymmetric coupling, resulting in undesired polarization or frequency shifts. 

\noindent However, horizontal or vertical displacement between the layers can modify the effective dielectric constant and affect the bandwidth and resonance frequency. And those effects are not all bad. They can be further utilized to fine tune the antenna. 

\noindent The article emphasizes that precise alignment is crucial, especially in multilayer configurations or designs that utilize proximity coupling, aperture coupling, or stacked patches, where even small displacements can significantly change the strength and nature of the coupling. Therefore, careful mechanical assembly and simulation-based tuning are essential to ensure consistent performance and to minimize coupling-related degradation.

\begin{figure}[H] 
    \centering 
    \includegraphics[width=1.0\textwidth]{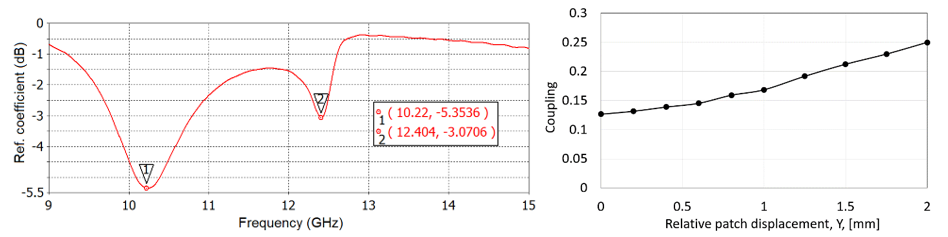} 
    \rule{35em}{0.5pt} 
    \caption {The example response of two-antenna system (left); Coupling as function of top patch displacement in the Y-axis (right)\cite{bulja2023microstrip}.}
    \label{fig:coupling} 
\end{figure}

\noindent The brief idea of this methodology can be expressed as the equation \ref{eq:coupling}. And its effect is shown in the figure \ref{fig:coupling}.

\begin{equation}
\left\{
\begin{aligned}
K(d) &= K_0 \cdot \alpha d \\
K_0 &= \frac{\Delta f}{f_{\text{mid}}}, \quad \text{where } \Delta f = f_1 - f_2,\quad f_{\text{mid}} = \frac{f_1 + f_2}{2}
\end{aligned}
\right.
    \label{eq:coupling}
\end{equation}

\vspace{1em}
\textbf{Where:}
\begin{itemize}
    \item $K(d)$: Coupling strength as a function of displacement $d$
    \item $K_0$: Coupling strength when layers are perfectly aligned (i.e., $d = 0$)
    \item $d$: Effective displacement between the layers (in mm)
    \item $\alpha$: Constant dependent on dielectric constant $\varepsilon_r$, substrate height $h$, and frequency $f$, etc.
    \item $f_1 \& f_2$: Resonant frequencies of two layers of patches
    \item $\Delta f$: Frequency difference between two layers
    \item $f_{\text{mid}}$: Center frequency, defined as $\frac{f_1 + f_2}{2}$
\end{itemize}

\subsubsection{Correction Procedure}
\begin{figure}[H] 
    \centering 
    \includegraphics[width=1.0\textwidth]{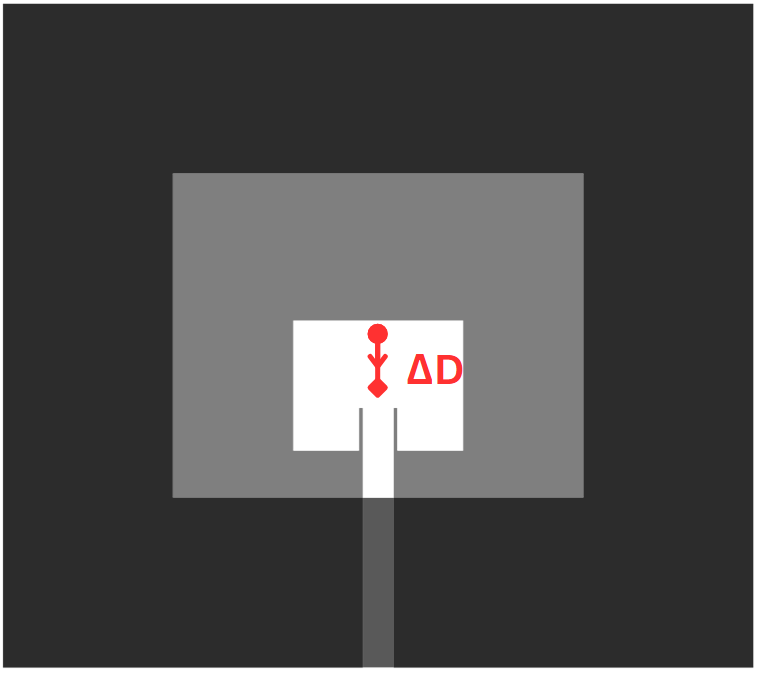} 
    \rule{35em}{0.5pt} 
    \caption {The 5 mm displacement introduced for the two layers of patches}
    \label{fig:displacement} 
\end{figure}

\noindent Taking the methodologies mentioned above into account. Displacement along the Y-axis, which was parallel to the feed line, was introduced to alter the resonant frequencies of the dual band stacked patch antenna as well as further fine tune the antenna.
\noindent After trying various different displacement parameter, a 5 mm displacement along the Y-axis was finally introduced as shown in the figure \ref{fig:displacement}. The square represents the center of the top patch, and the dot represents the center of the middle layer patch. 
\subsubsection{Corrected Results}
\begin{figure}
     \centering
     \begin{subfigure}[b]{0.45\textwidth}
         \centering
         \includegraphics[width=\textwidth]{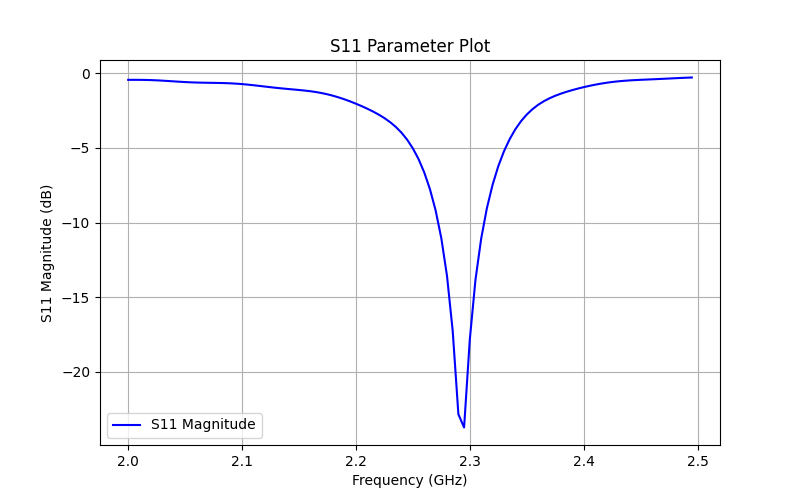}
         \caption{lower band S11}         
     \end{subfigure}
     \hfill
     \begin{subfigure}[b]{0.45\textwidth}
         \centering
         \includegraphics[width=\textwidth]{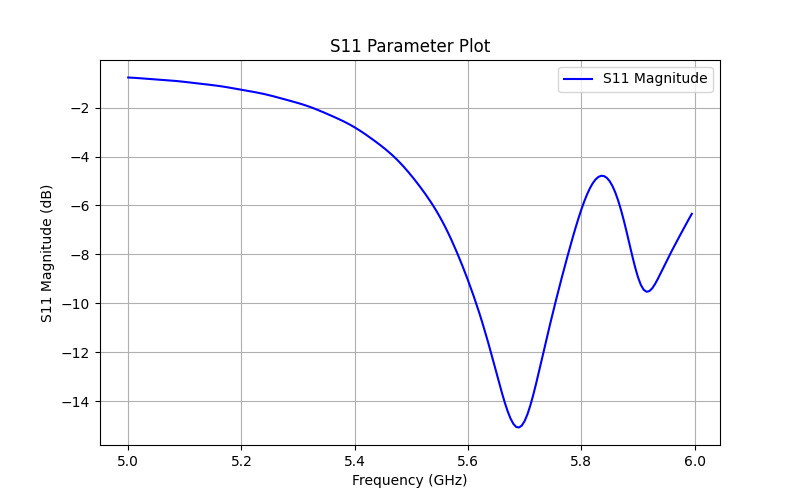}
         \caption{upper band S11}
     \end{subfigure}
        \caption{The corrected S11 simulation results}
        \label{fig:s11_refine}
\end{figure}

\begin{figure}
     \centering
     \begin{subfigure}[b]{0.45\textwidth}
         \centering
         \includegraphics[width=\textwidth]{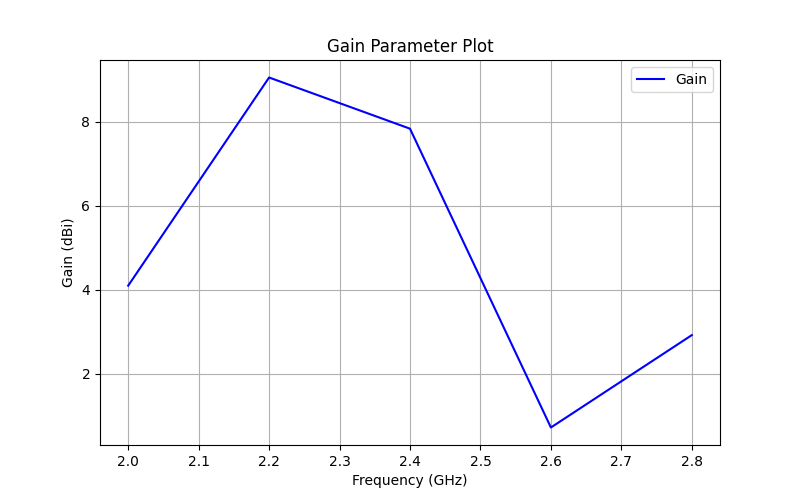}
         \caption{lower band Gain}         
     \end{subfigure}
     \hfill
     \begin{subfigure}[b]{0.45\textwidth}
         \centering
         \includegraphics[width=\textwidth]{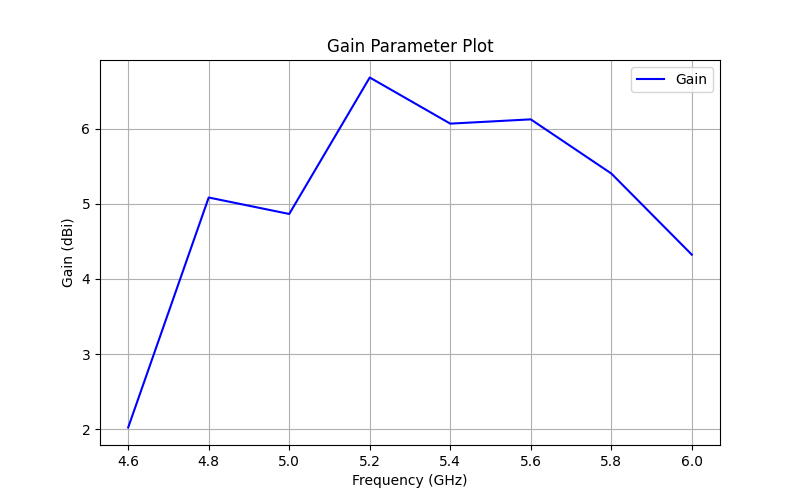}
         \caption{upper band Gain}
     \end{subfigure}
        \caption{The corrected Gain simulation results}
        \label{fig:gain_refine}
\end{figure}

\begin{figure}
     \centering
     \begin{subfigure}[b]{0.45\textwidth}
         \centering
         \includegraphics[width=\textwidth]{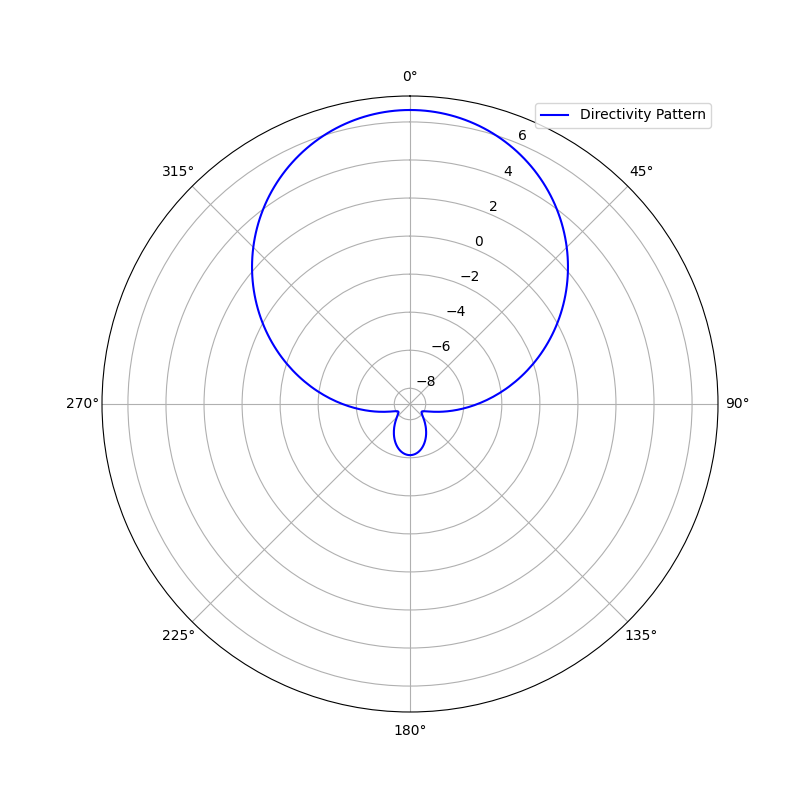}
         \caption{lower band farfield}         
     \end{subfigure}
     \hfill
     \begin{subfigure}[b]{0.45\textwidth}
         \centering
         \includegraphics[width=\textwidth]{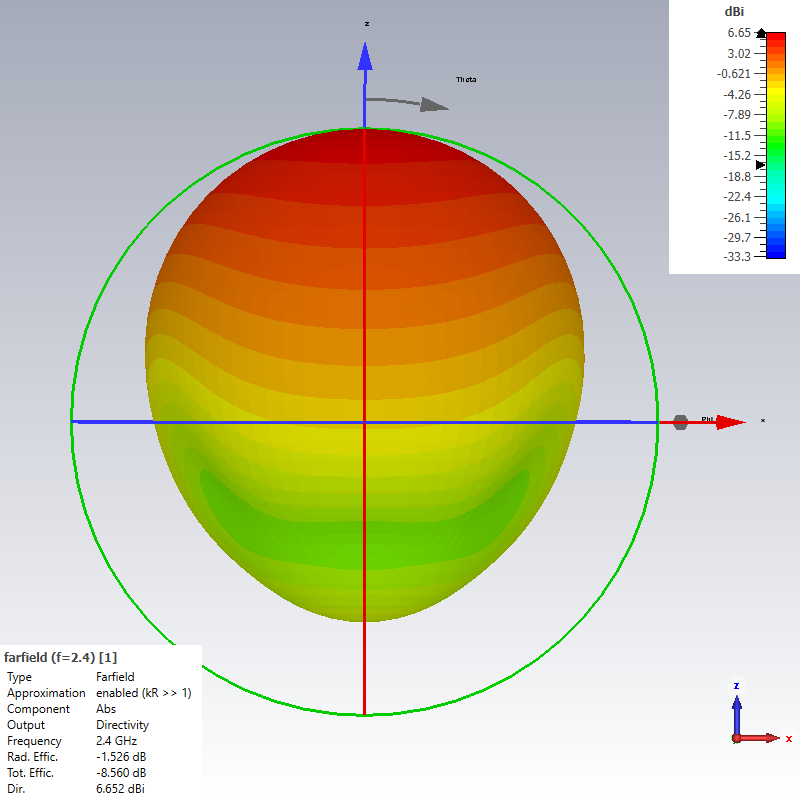}
         \caption{lower band 3D farfield}
     \end{subfigure}
        \caption{The corrected lower band farfield simulation results}
        \label{fig:lower_farfield_refine}
\end{figure}

\begin{figure}
     \centering
     \begin{subfigure}[b]{0.45\textwidth}
         \centering
         \includegraphics[width=\textwidth]{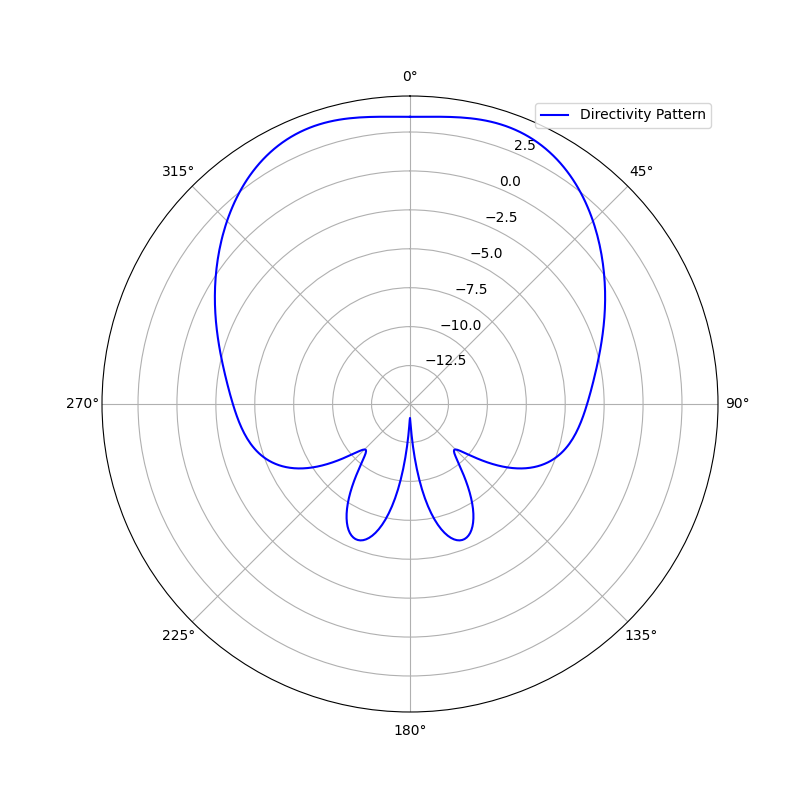}
         \caption{upper band farfield}         
     \end{subfigure}
     \hfill
     \begin{subfigure}[b]{0.45\textwidth}
         \centering
         \includegraphics[width=\textwidth]{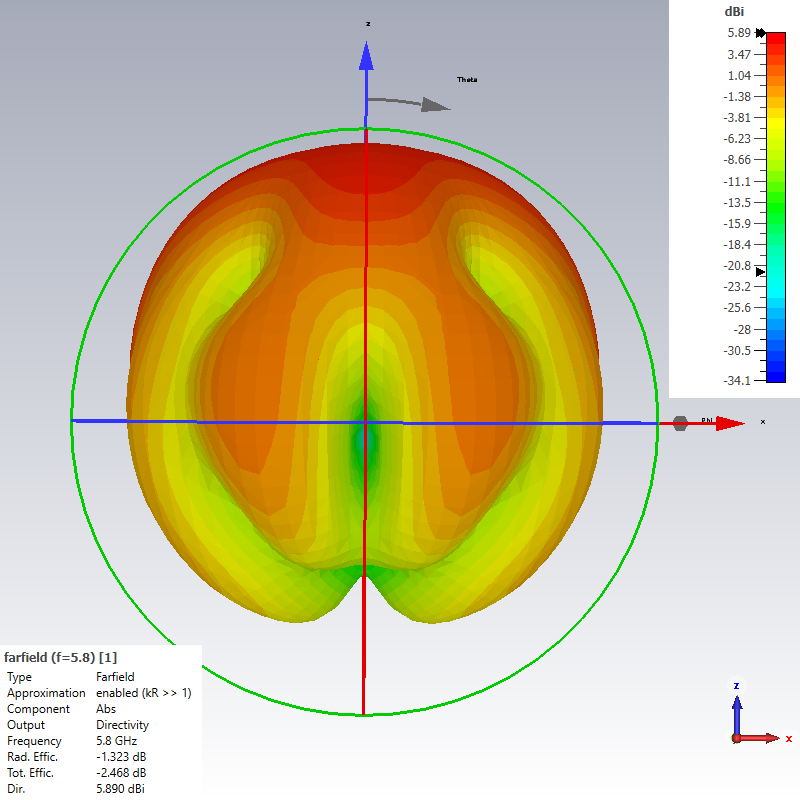}
         \caption{upper band 3D farfield}
     \end{subfigure}
        \caption{The corrected upper band farfield simulation results}
        \label{fig:upper_farfield_refine}
\end{figure}

\noindent After the corrected patch antenna was simulated in the CST software, there was a huge improvement with regard to the performance. The figure \ref{fig:s11_refine} shows the S11 simulation results. The figure \ref{fig:gain_refine} shows the gain results. The figure \ref{fig:lower_farfield_refine} and \ref{fig:upper_farfield_refine} shows the farfield results.

\noindent It can be observed from the figures that this edition of the stacked dual band patch antenna reached a pretty feasible performance with S11 in two bands low enough and with overall wide bandwidth. the Gain was fairly stable across the spectrum. the farfield plots demonstrated that the antenna was equipped with high directivity. Moreover, the radiation efficiency in both resonant frequencies were sitting between -1.3 to -1.5dB, which transformed to the percentage format, meant that at least 70\% of the power injected was successfully radiated.

\section{Fabrication of the Patch Antenna}
\subsection{Export of the Design for Mask Printing}
\noindent Since all the structure and the parameter of the dimension were finally determined, the project progressed to the fabrication phase, which required a lot of communication and coordination among various of departments in the university.

\noindent In order to fabricate the design to the substrates, the 3D modeling of the patch antenna needs to be transformed to the 2D CAD design and print on the mask. Cross-cut surfaces were chosen through the copper layer to show what was the etching pattern on the substrates. Then some cross marks were added to ensure the alignment during the etching process.

\begin{figure}[H] 
    \centering 
    \includegraphics[width=1.0\textwidth]{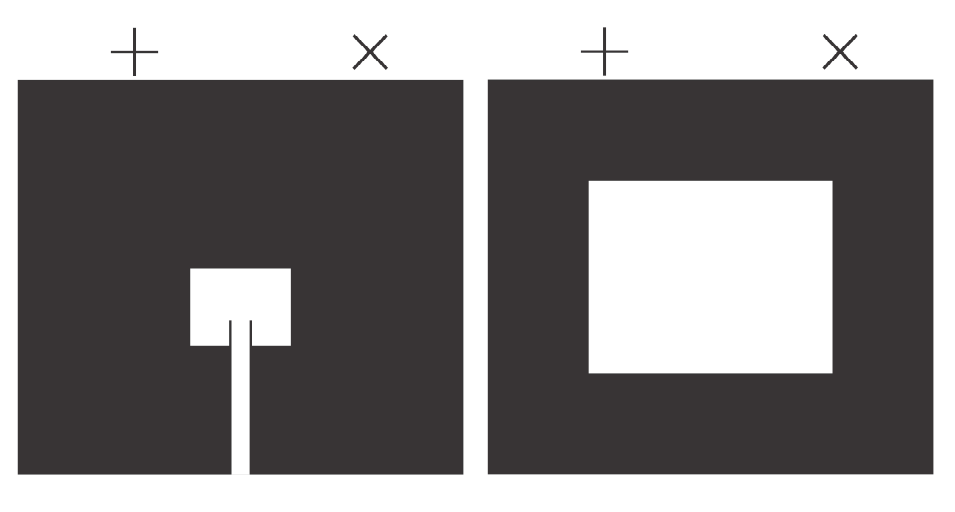} 
    \rule{35em}{0.5pt} 
    \caption {The fabrication masks for top copper}
    \label{fig:fab_top} 
\end{figure}

\begin{figure}[H] 
    \centering 
    \includegraphics[width=1.0\textwidth]{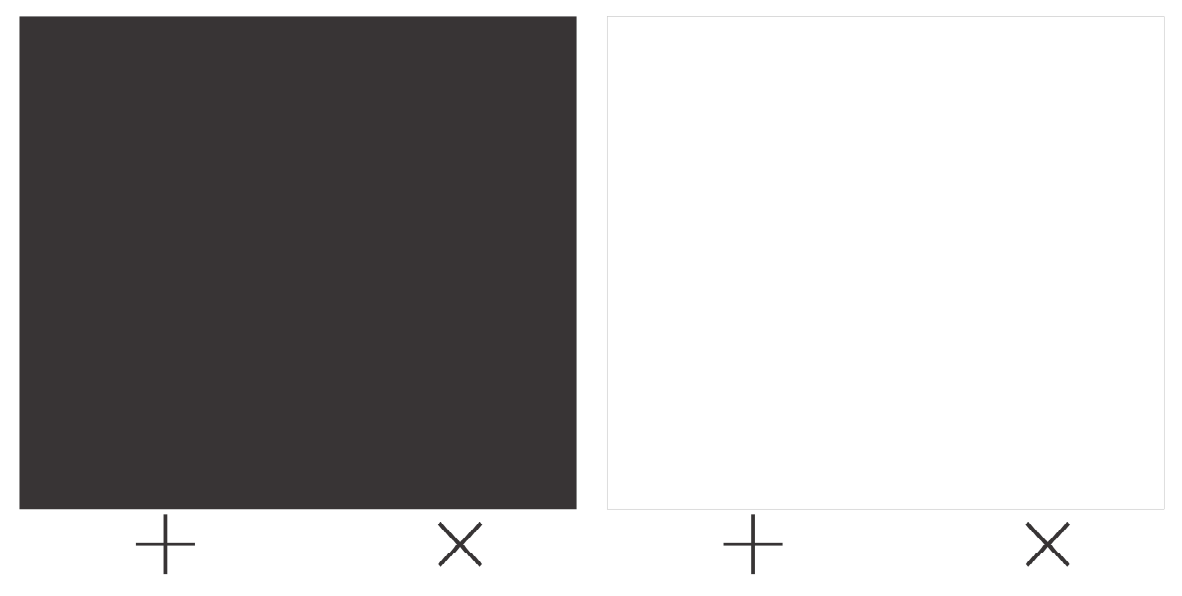} 
    \rule{35em}{0.5pt} 
    \caption {The fabrication masks for bottom copper}
    \label{fig:fab_bottom} 
\end{figure}

\noindent After all the CAD works were done, the generated masks were printed on a transparent film, as what is shown in the figure \ref{fig:fab_top} for the patches and figure \ref{fig:fab_bottom} for the ground.

\subsection{Fabrication on the Substrate and SMA Connection}
\begin{figure}[H] 
    \centering 
    \includegraphics[width=1.0\textwidth]{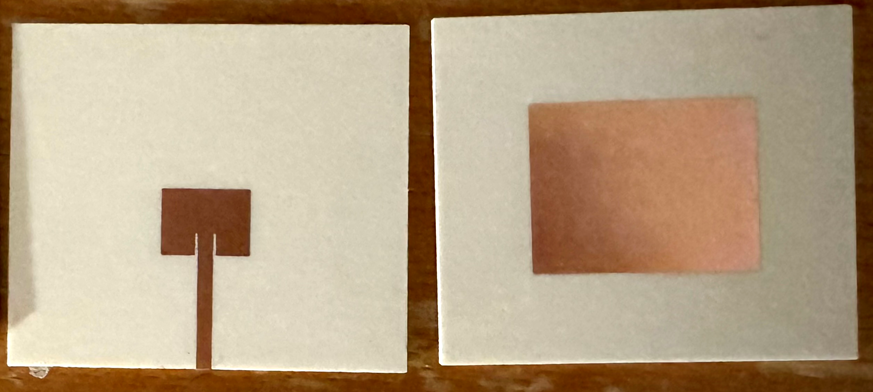} 
    \rule{35em}{0.5pt} 
    \caption {The fabricated patches}
    \label{fig:fab_out} 
\end{figure}

\noindent When the masks were printed, they were handed over to the electronics workshop in the university along with a SMA connector. the two layers of patches were firstly fabricated as shown in the figure \ref{fig:fab_out}. Then the two patches were stacked according to the structure preciously shown in the figure \ref{fig:stack_3d}, and a SMA connector was soldered to the feed line to provide the antenna with universal connection. 

\begin{figure}[H] 
    \centering 
    \includegraphics[width=0.8\textwidth]{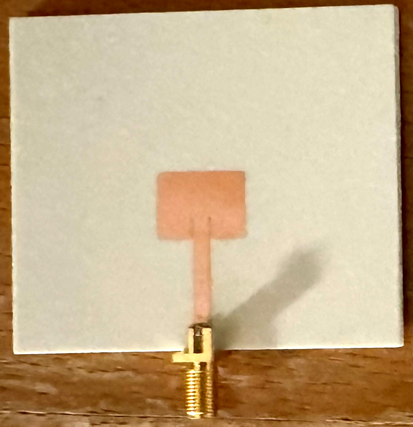} 
    \rule{35em}{0.5pt} 
    \caption {The fabricated patches}
    \label{fig:fab_sma} 
    
\end{figure}

\noindent At last, the dual band patch antenna with stacked structure was finally finished, as shown in the figure \ref{fig:fab_sma}.

\chapter{Conclusion}
\chaptermark{Conclusion}

{
\hypersetup{linkcolor=black}
\minitoc
}

\section{Problems Haven't Anticipated at the Beginning}
\noindent Several unforeseen challenges emerged during the progression of this project, highlighting areas that could have been better planned or anticipated. Initially, too much emphasis was placed on conducting broad background research without a clear focus on the final implementation. While understanding the theoretical foundation is essential, an excessive amount of time was spent reviewing general literature and existing works, which in hindsight could have been more productively allocated to early conceptualization and simulation of novel designs. 

\noindent Conversely, the project also suffered from insufficiently deep research when it came to selecting the design methodology. Rather than decisively choosing between the slot-based and stacked patch antenna designs based on a comprehensive feasibility analysis, both paths were explored in parallel. This dual-track approach diluted time and resources, ultimately leading to delays and inefficiencies. 

\noindent Additionally, the fabrication process introduced unexpected logistical hurdles. The process of coordinating with multiple departments within the university—such as the electronics workshop and technical support staff—proved to be far more time-consuming than initially anticipated. A more efficient approach would have been to finalize fabrication details and interdepartmental communication protocols earlier in the timeline. 

\noindent Furthermore, the lack of access to advanced measurement equipment, such as a vector network analyzer (VNA) and anechoic chamber, significantly limited the empirical evaluation of the antenna’s performance. In retrospect, proactively seeking assistance from a postgraduate student or a researcher with access to such facilities could have provided essential experimental validation and accelerated the iteration process. These lessons underscore the importance of strategic planning, timely decision-making, and proactive resource acquisition in engineering design projects.

\section{Limitations}
\noindent When it comes to the limitation of this project, or to the patch antenna it self,despite demonstrating dual-band functionality, several key limitations should be acknowledged. First, the bandwidth of the proposed patch antenna is relatively narrow. This is an inherent drawback for most patch antennas due to their fundamental resonant nature and the constraints imposed by the thin dielectric substrates; while techniques such as stacking or adding slots can widen the bandwidth, substantial increases are typically challenging without compromising other performance metrics. 

\noindent Second, the design in its current form is not perfectly compatible with standard SMA connectors: the feed structure needs to be rotated by 90 degrees to accommodate the thick, multi-layer substrate, which in turn introduces a slight asymmetry and may influence the radiation pattern.

\noindent Third, the fine-tuning work has been limited by time constraints. Although initial dimensions were derived from an online calculator, no in-depth iterative optimization was conducted, meaning the 5 mm displacement used to adjust coupling between the top and middle patches may not be optimal. In practice, small refinements of the patch dimensions, interlayer spacing, and feed position could significantly improve impedance matching, bandwidth, and overall efficiency.

\noindent Finally, there is currently a lack of thorough measurement data to validate and refine the design. The bachelor-level project had no access to a vector network analyzer (VNA) for precise S-parameter measurements, nor to an anechoic chamber for far-field pattern characterization. Without these essential experimental evaluations, certain critical aspects of performance—such as real-world bandwidth, gain, and radiation efficiency—remain partly unverified, limiting the practical assessment of this dual-band patch antenna.
\section{Potential Application}
\noindent The ultra-thin profile and planar structure of the designed patch antenna render it highly suitable for a diverse array of modern wireless applications where compactness, low weight, and mechanical robustness are paramount. In particular, its minimal thickness makes it an excellent candidate for integration into Wireless Body Area Networks, where conformal, low-profile antennas are essential for unobtrusive on-body communication systems.

\noindent Additionally, the antenna's lightweight and compact design are advantageous for aerospace applications, including aircraft and spacecraft, where reducing aerodynamic drag and conserving space are critical considerations. The antenna's directional radiation characteristics also make it well-suited for Global Positioning System (GPS) devices, facilitating accurate positioning and navigation.

\noindent Furthermore, its applicability extends to Radio Frequency Identification (RFID) systems, where efficient signal transmission and reception are vital for object tracking and data communication. In the realm of wireless communications, the antenna can be effectively utilized in Wi-Fi routers and mobile devices, supporting seamless connectivity in both indoor and outdoor environments. Its potential use in 5G applications is also notable, where high-frequency operation and compact form factors are increasingly demanded. Moreover, the antenna's suitability for integration into printed circuit boards makes it an ideal choice for various consumer electronics, enabling streamlined manufacturing processes and cost-effective production. 

\noindent Overall, the designed patch antenna's characteristics align well with the requirements of contemporary wireless systems across multiple sectors, including healthcare, aerospace, telecommunications, and consumer electronics.

\section{Suggested Further Modifications and Development}
\noindent To enhance the performance and practicality of the dual-band patch antenna, several modifications and future development directions are recommended.

\noindent First and foremost, further optimization of the antenna geometry through parametric tuning is necessary. The current design relies on initial dimension estimations from online calculators and a fixed 5 mm displacement for patch alignment, which may not yield the optimal coupling or impedance matching. A more rigorous approach involving simulation-based iterative optimization could significantly improve bandwidth, gain, and resonance accuracy.

\noindent Additionally, exploring alternative feeding mechanisms, such as proximity or aperture coupling, may enhance impedance matching and reduce feed-line losses, especially for higher frequency bands. From a structural perspective, using advanced substrate materials with lower dielectric loss and higher permittivity could help improve radiation efficiency and reduce the overall antenna footprint. The current configuration may also benefit from mechanical refinement; for instance, the SMA connector interface could be redesigned to ensure compatibility with standard connector orientations, eliminating the need for a 90-degree adjustment that disrupts symmetry.

\noindent Experimentally, the antenna should undergo comprehensive testing using a vector network analyzer (VNA) and anechoic chamber to validate its performance in real-world conditions. This data can inform further iterations and corrections. Moreover, integrating the antenna into a flexible or conformal substrate could expand its applicability in wearable and body-mounted systems. Future designs may also consider extending the functionality beyond dual-band operation, exploring tri-band or ultra-wideband designs to support more complex communication systems.

\noindent Finally, transitioning from a discrete prototype to an integrated system by embedding the antenna directly into printed circuit boards (PCBs) for specific applications, such as IoT or WBAN devices, could improve manufacturability and commercial viability. These improvements, combined with empirical validation and advanced modeling, would significantly elevate the design's maturity and readiness for deployment.



\label{Bibliography}
\printbibliography[title={References}, heading=bibintoc, resetnumbers=true]
\begin{refsection}
\DeclareFieldFormat{labelnumberwidth}{#1}
\nocite{*}
\printbibliography[title={Bibliography}, notcategory=cited, omitnumbers=true, heading=bibintoc]
\end{refsection}

\end{document}